\documentclass[conference]{IEEEtran}
\IEEEoverridecommandlockouts
\usepackage{cite}
\usepackage{calligra}
\usepackage{amsmath,amssymb,amsfonts}
\usepackage{algorithmic}
\usepackage{graphicx}
\usepackage{textcomp}
\usepackage{xcolor}
\usepackage{float}
\usepackage{mathtools}
\usepackage{makecell}
\usepackage{subfigure}
\usepackage{soul}
\usepackage{multirow,color,todonotes}
\usepackage{array} 
\usepackage{tikz,balance}
\usepackage{url}
\usepackage{enumitem,xcolor,booktabs,colortbl}
\usepackage[linesnumbered, ruled]{algorithm2e}
\SetKwRepeat{Do}{do}{while}
\def\BibTeX{{\rm B\kern-.05em{\sc i\kern-.025em b}\kern-.08em
    T\kern-.1667em\lower.7ex\hbox{E}\kern-.125emX}}

\usepackage[letterpaper, left=0.67in, right=0.67in, top=0.75in, bottom=1.05in]{geometry}

\begin{document}
\title{\LARGE \bf PsybORG$^+$: Modeling and Simulation for Detecting Cognitive Biases in Advanced Persistent Threats\vspace{-3mm}} 

 \author{
\IEEEauthorblockN{Shuo Huang\IEEEauthorrefmark{1}, Fred Jones\IEEEauthorrefmark{2}, Nikolos Gurney\IEEEauthorrefmark{3}, David  Pynadath\IEEEauthorrefmark{3}, \\ Kunal Srivastava\IEEEauthorrefmark{2},  Stoney Trent\IEEEauthorrefmark{4},  Peggy Wu\IEEEauthorrefmark{2}, 
Quanyan Zhu\IEEEauthorrefmark{1}}
\IEEEauthorblockA{\IEEEauthorrefmark{1}\textit{Department of Electrical and Computer Engineering}, \\ \textit{New York University, New York 10012, USA} 
\{sh7467, qz494\}@nyu.edu}
\IEEEauthorblockA{\IEEEauthorrefmark{2}\textit{Raytheon Technologies, USA} \quad
\{Frederick.Jones, Kunal.Srivastava, Peggy.Wu\}@rtx.com}
\IEEEauthorblockA{\IEEEauthorrefmark{3}\textit{University of Southern California} \quad
\{gurney,pynadath\}@ict.usc.edu}
\IEEEauthorblockA{\IEEEauthorrefmark{4}\textit{Bulls Run Group} \quad stoney@bullsrungroup.com}
\vspace{-10mm}
}

\IEEEoverridecommandlockouts
\maketitle

\begin{abstract}






Advanced Persistent Threats (APTs) bring significant challenges to cybersecurity due to their sophisticated and stealthy nature. Traditional cybersecurity measures fail to defend against APTs. Cognitive vulnerabilities can significantly influence attackers' decision-making processes, which presents an opportunity for defenders to exploit. This work introduces PsybORG$^+$, a multi-agent cybersecurity simulation environment designed to model APT behaviors influenced by cognitive vulnerabilities. A classification model is built for cognitive vulnerability inference and a simulator is designed for synthetic data generation. 
Results show that PsybORG$^+$ can effectively model APT attackers with different loss aversion and confirmation bias levels. The classification model has at least a 0.83 accuracy rate in predicting cognitive vulnerabilities.
\end{abstract}


\section{Introduction}

In recent years, Advanced Persistent Threats (APTs) have become one of the most serious challenges in cybersecurity. These attacks are characterized by their sophisticated, stealthy nature and are often carried out by well-resourced adversaries \cite{chen2014study}. According to records in MITRE ATT\&CK\cite{strom2018mitre}, APTs' tactics and techniques are becoming increasingly complex and advanced. Traditional cybersecurity measures have proven insufficient in defending against the growing threat posed by APTs\cite{fahad2023securing}. It is necessary to design more advanced and proactive defense mechanisms.

Cognitive vulnerabilities, or biases, can widely affect our judgments and decisions in daily life. In cybersecurity, attackers with different cognitive vulnerabilities display significantly different behaviors. For example, the attacker with sunk cost fallacy spends more time applying resources that they have invested in. It is important to identify and exploit the cognitive vulnerabilities of potential APT attackers. 


To simulate the behaviors of APT attackers influenced by various cognitive vulnerabilities, we develop a multi-agent cybersecurity simulation environment called PsybORG$^+$, which models APTs as a Hidden Markov Model (HMM). We also build a classification model to do cognitive vulnerability inference and a simulator for synthetic data generation. 

We test our model on an artificial dataset. The results show that the classification model has at least 0.83 accuracy rate in the prediction of 3 cognitive vulnerabilities. We compare the simulation results from our simulator with those generated by using real and random parameters. We find that the average distance between our synthetic data and the real parameters' results was small for loss aversion and confirmation bias actions. However, The there parameters' performance in the simulation of attackers with sunk cost fallacy are similar, which means PsybORG$^+$ is less effective in modeling sunk cost fallacy. 


\vspace{-3mm}\section{Related Work}
Modeling APTs requires an understanding of their life cycle. The MITRE ATT\&CK framework, a comprehensive knowledge base of cyber threat tactics and techniques \cite{strom2018mitre}, categorizes APT behaviors into 14 distinct tactics. APTs with different objectives leverage various combinations of these tactics. Many studies, including \cite{zhu2018multi,huang2018analysis,zhu2013game,huang2020dynamic}, have modeled APT attacks using this multi-stage, multi-phase structure.
The detection of APTs is challenging due to their stealthy, sophisticated, and persistent nature. Provenance Graph Analysis is a widely used technique for the detection of APTs \cite{han2020unicorn}. This method constructs a directed cyclic graph to model interactions in the network and analyzes the graph to detect anomalous behaviors associated with APTs. Machine learning is also applied in APT detection. Models trained on various network data can identify patterns and anomalous behaviors indicative of APTs. In \cite{alrehaili2021hybrid}, the authors developed the SAE-LSTM and CNN-LSTM models to detect signs of APTs. In \cite{eke2019use}, the authors utilized the LSTM-RNN model for APT detection. In \cite{joloudari2020early}, the C5.0 decision tree and Bayesian network were employed to detect and classify APTs using the NSL-KDD dataset.


Cognitive vulnerability is a psychological concept which has received increaing attention in cybersecurity. Seminal work in \cite{kahneman1984choices, tversky1981framing, tversky1988contingent} found that preferences can significantly influence decision-making processes. The authors of \cite{lemay2018cognitive} studied the influence of base rate fallacy, confirmation bias, and hindsight bias on APTs. In the recent study \cite{ferguson2021examining}, it examined the psychology of perception, decision-making, and behavior in the context of cyber attacks. Specifically, it investigated how attackers (red teamers) respond to defensive deception tactics, both cyber and psychological, within a controlled environment. 



\section{Preliminary}\label{prelim}
APT attackers have several cognitive vulnerabilities that defenders can exploit, such as base rate neglect, confirmation bias, loss aversion, and the sunk cost fallacy. This section introduces the behavioral models of these biases, which will be incorporated into  PsybORG$^+$  for analysis and simulation.

\subsection{Base rate neglect}
\begin{figure}[htb]
    \centering
 \vspace{-4mm}   \includegraphics[width=.38\textwidth]{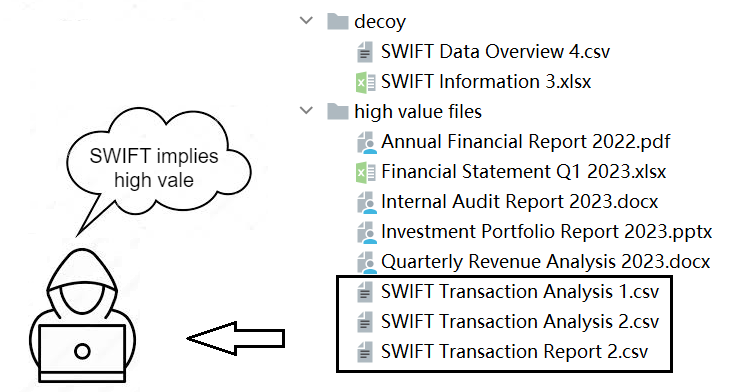}\vspace{-3mm}
    \caption{Base rate neglect of an APT attacker. In the attacker's view (black rectangle), all high-value files have 'SWIFT' in their filenames. This attacker may hold a belief that 'SWIFT' implies high value. However, these files only account for a small portion of the high-value files. To exploit this cognitive bias, the defender can deploy decoy files containing 'SWIFT' in their filenames to attract the attacker.}
    \label{brn_example}\vspace{-5mm}
\end{figure}

Base rate neglect is a cognitive bias where individuals tend to overweight the representativeness of a piece of evidence while ignoring its base rate, or how often it occurs\cite{kahneman1973psychology}. In cybersecurity, this bias can affect APT attackers, leading them to make more attempts on filenames or account names that sound significant. For instance, if an APT attacker exhibits base rate neglect and encounters a specific keyword in the filenames of high-value files, they might erroneously believe that the presence of this keyword consistently indicates high value, as illustrated in Figure \ref{brn_example}.

\subsection{Confirmation bias}

Confirmation bias is the tendency to overweight confirming evidence \cite{nickerson1998confirmation}. In cybersecurity, this bias can be observed in an attacker's behavior, particularly in the time spent confirming the reliability of their hypotheses. For instance, if an APT attacker finds a credential file for a server, he may hypothesize that the server exists and contains important files. Even after many failed login attempts, the attacker might not abandon this hypothesis, believing that the server exists but has not yet been found. This persistence, driven by confirmation bias, illustrates the difficulty of falsifying a hypothesis once it has been formed.

Assume that $\lambda_c \in [0,1]$ is the rate of finding confirming evidence within all credential file checking actions. If $\lambda_c$ is significantly greater than 0.5, we can say this attacker has a high confirmation bias.

\subsection{Loss aversion}
Loss aversion refers to a cognitive vulnerability leading to a negative emotional reaction to losses, even facing more gains\cite{schmidt2005loss}. APT attackers with loss aversion prefer to take low-risk measures to gather information. These attackers only scan the most common ports rather than all common ports at the initial stage of service discovery. Then, it would stealthily scan other ports. As this activity resembles normal network behaviors, these attackers are less likely to alert the defender.

According to prospect theory\cite{kahneman1979prospect}, attackers' asymmetric perceptions of loss or gain $\omega \in \mathbb{R}$ can be represented by the subjective utility function $\epsilon(\omega,\lambda_l)$, in which $\lambda_l \in \mathbb{R}_+$ denotes the coefficient controlling the loss aversion.

In the service discovery process, the loss aversion can be modeled as (1)-(2). $a \in \mathbb{R}$ and $s \in \mathbb{R}$ represent the estimated loss or gain of aggressive service discovery and stealth service discovery respectively. $\gamma(a,s,\lambda_l) \in [0,1]$ is the probability of taking aggressive service discovery. $\rho \in \mathbb{R}$ represents the parameter controlling the curvature of $\epsilon(\omega,\lambda_l)$. $\mu \in \mathbb{R}$ is the logit sensitivity, which is used to adjust the stability of the decision-making process. 
\begin{equation}
\gamma(a,s,\lambda_l) := \frac{1}{1 + e^{-\mu(\epsilon(a,\lambda_l)-\epsilon(s,\lambda_l))}}
\end{equation}
\begin{equation}
\epsilon(\omega,\lambda_l) :=
\begin{cases} 
    \omega^\rho & \text{if } \omega \geq 0 \\
    -\lambda_l (-\omega)^\rho & \text{if } \omega < 0 
\end{cases}
\end{equation}


\subsection{Sunk cost fallacy}
\begin{figure}[htb]
    \centering
  \vspace{-3mm}  \includegraphics[width=.33\textwidth]{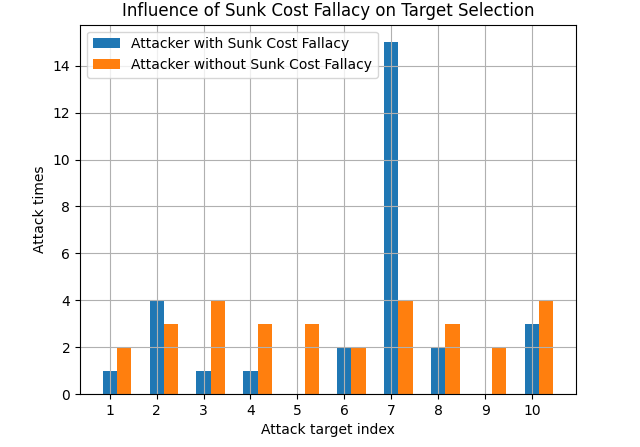}
   \vspace{-2mm} \caption{Influence of sunk cost fallacy: The figure compares attack patterns of attackers with (blue) and without (orange) sunk cost fallacy across different targets. The attacker influenced by sunk cost fallacy shows a strong preference for target 7, investing significantly more attempts (about 15) compared to other targets. This behavior reflects the tendency to persist with a chosen path due to previous investment.}
    \label{sunk cost fallacy example}\vspace{-2mm}
\end{figure}

The sunk cost fallacy describes the tendency to make irrational decisions due to previously invested resources\cite{friedman2007searching}. APT attackers with sunk cost fallacy prefer to spend time and resources on exploits they have invested in.
For example, 
An attacker targets an encrypted file, File X, and invests resources in attempts to decrypt it. Despite facing many obstacles, this attacker continues to crack File X, as shown in Figure \ref{sunk cost fallacy example}.

Suppose that there are $Z$ target files or servers available for exploiting, the 
perceived value of a target $z \in \mathcal{Z}:= \{1,...,Z\}$ can be modeled by a function $L(z): \mathcal{Z} \rightarrow \mathbb{R}$. (3) shows a linear model of $L(z)$, in which $r(z): \mathcal{Z} \rightarrow \mathbb{R}$ is the estimated reward function for investing resource on $z$, $c(z): \mathcal{Z} \rightarrow \mathbb{R}$ is the function of sunk cost spent on $z$. $\lambda_s \in \mathbb{N}_+$ is coefficient controlling the sunk cost fallacy. The probability of choosing target $z$ is presented in (4).
\begin{eqnarray}
& & L(z) := r(z) + \lambda_s c(z)\\
& & p_s(z) := \frac{L(z)}{\Sigma_{j=1}^{Z} L(j)}
\end{eqnarray}

\section{Advanced Persistent Threat Modeling}
This section presents an integrative model that combines APT threat behaviors with human cognitive biases. This integrative modeling is the backbone of the PsybORG$^+$ framework. It allows for behavior-driven inference of cognitive biases and facilitates simulation and data generation. The three cognitive biases introduced in Section \ref{prelim} will be incorporated into PsybORG$^+$  as a case study to demonstrate its capabilities.
\subsection{APT hidden Markov model}

Consider an APT attacker that has $N \in \mathbb{N}_+$ biases. Each bias is characterized by a set of types $V_n, n=1, \cdots, N$.  Bias $n$ of type $v_n\in V_n$ is characterized by the associated parameter $\lambda_{v_n}\in \Lambda_{v_n}$, where $\Lambda_{v_n}$ is the set of values the parameter can take. For example, the loss aversion bias of the attacker can take different levels, e.g., high or low; hence type $v_l\in V_l:=\{\theta_{HL}, \theta_{LL}\}$, where $l\in \{1, 2, \cdots, N\}$ is the index associated with loss aversion, and $\theta_{HL}$ refers to the type of high loss aversion and $\theta_{LL}$ refers to the type of low loss aversion.
The cognitive bias state of the attacker is vector $\theta=\{v_n\}_{n=1}^N, v_n\in V_n, n=1, \cdots, N$. The state attribute is thus characterized by the vector $\lambda=\{\lambda_{v_n}\}_{v_n\in\theta} \in \Lambda:=\prod_{n=1}^n\prod_{v_n\in V_n}\Lambda_{v_n}$. Let $\Theta:=\prod_{n=1}^nV_n$ be the set of all possible cognitive states. 
For each bias state $\theta \in \Theta$, a distribution $p(\lambda|\theta)$ is used to characterize the certainties at each state. Let $\lambda\in\Lambda$ be interpreted as the factors that influence the bias state. A sample from the distribution determines the attribute of a given bias state $\theta$. 

A bias state $\theta\in\Theta$ determines the attack behavior which can be modeled through the transition of cyber states. To this end, we first define $\mathcal{Q}:= \{K, S, U, R\}$ as the set of cyber stages describing the APT life cycle. Each cyber stage $q \in \mathcal{Q}$ represents the attacker's levels of knowledge and privilege of a host, as depicted in Table \ref{state_in_fsm}. {The cyber state space is not confined to the sample baseline set 
$\mathcal{Q}$. Generally, a more detailed cyber state space 
$\mathcal{X}$ can capture finer-grained steps in the cyber kill chain compared to the baseline state space 
$\mathcal{Q}$, where $\mathcal{Q}\subseteq \mathcal{X}$. }

\vspace{-3mm}\begin{table}[htb]
\centering
\caption{\label{state_in_fsm}APT cyber stages}
\begin{tabular}{|c|c|}
	\hline Stage&Description\\
	\hline K&The host’s IP address is known.\\
        \hline S&The host's services are known.\\
        \hline U&The attacker has a user shell on the host.\\
        \hline R&The attacker has a root shell on the host.\\
    \hline
\end{tabular}\vspace{-2mm}
\end{table}

Considering the potential dependency among some attack behaviors, we model an APT attacker as a probabilistic finite state machine (PFSM). We define $\mathcal{A}:=\{\mathcal{A}_q\}_{q \in \mathcal{Q}}$ as the action space, where $\mathcal{A}_q$ is the action set available for an APT attacker in stage $q$ and $\mathcal{A}_q \cap \mathcal{A}_{q'} = \varnothing$ for $q \neq q'$. We also define $\mathcal{H}:=\Theta \times \mathcal{Q}$ as the state space. 

\begin{figure}[htb]
    \centering
    \vspace{-4mm}\includegraphics[width=.35\textwidth]{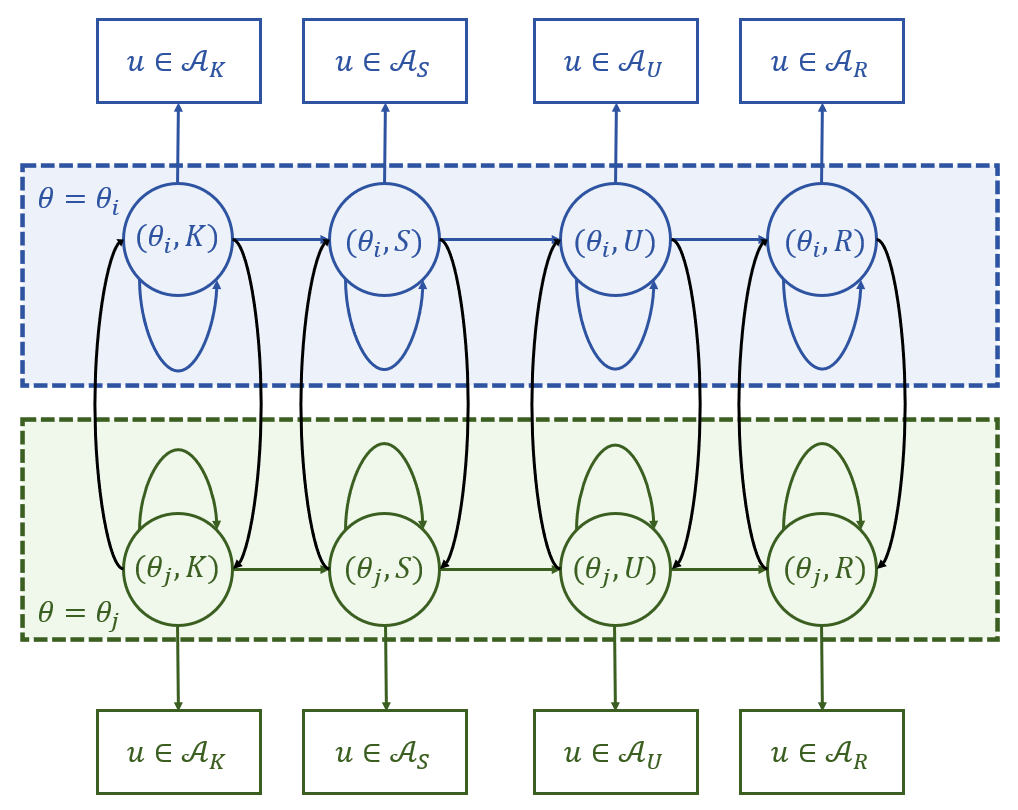}
    \caption{An APT Hidden Markov Model: Attackers can take actions (blue and green  lines) to transition between different life cycle stages. Each dotted box represents an APT attack life cycle. The transition between biases (black lines) happens if and only if the attacker is exposed to a trigger.}
    \label{bias_model}\vspace{-5mm}
\end{figure}

The integrated cyber and cognitive bias state is the joint state $y=(\theta, x)$, where $\theta\in\Theta$ is the cognitive bias state and $x\in X$ is the cognitive bias state; $Y=(\Theta, X)$ determines the state space of the HMM. At each state $y\in Y $, an attack action is observed with the kernel $p(\cdot|y)$. Let $u$ denote the action observed at the state $y$, which is determined by the cyber component of the joint state. Figure \ref{bias_model} depicts an example of the HMM with $X=Q$. In this case, at a given state $y\in Y$, the action $u\in \mathcal{A}_q$, where $q\in Q$. The HMM evolves over time. We use subscript $t$ to denote the state and the action at time $t$.

\subsection{Model driven biases inference}
We aim to infer the attackers' biases to help the defender design appropriate defensive strategies. We assume that $\mathcal{A}^*$ consists of all action sequences of the form $\mathbf{u}^l$=$\{u_1, \cdots, u_l\}$, where each $u_t \in \mathcal{A}$ for $t \in [1,\cdots,l]$ and $l \in \mathbb{N}_+$ is the length of the action sequence. Since the action sets in different cyber stages are disjoint, the
cyber stage is known if an action is given. We can maximize a posterior $p(\theta|\mathbf{u}^l)$  to find the biases $\theta \in \Theta$, which most likely generates a given action sequence $\mathbf{u}^l \in \mathcal{A}^*$. Our target can be represented as the following equations:
\begin{equation}
  \arg\max\limits_{\theta \in \Theta} \ p(\theta|\mathbf{u}^l)
\end{equation}   \vspace{-2mm} 
\begin{equation}
p(\theta|\mathbf{u}^t)=\frac{p(u_t|\theta)p(\theta|\mathbf{u}^{t-1})}{p(\mathbf{u}^t)}
\end{equation}  \vspace{-2mm} 
\begin{equation}
    p(u_t|\theta)=\int p(u_t|\lambda)p(\lambda|\theta) \, d\lambda
\end{equation}

This can be solved by the Bayesian inference algorithm if the initial distribution of biases $p(\theta)$ is given and $p(u_t|\theta)$ is computable for each $u_t \in \mathcal{A}$.

\subsection{Data driven biases inference}
Given that $p(\theta)$ and $p(u_t|\theta)$ are often unknown, we can only use action sequences to do nonparametric 
density estimation on $p(u_t|\theta)$. It is straightforward to compute the relative frequency for each possible choice of $u_t$ among action sequences generated by an attacker with bias state $\theta$. Then, we use the decision tree or neuron network to find $p(\theta|\mathbf{u}^l)$.

\subsection{PsybORG$^+$}
We develop a multi-agent cybersecurity simulation environment called PsybORG$^+$ to simulate the behaviors of APT attackers influenced by various cognitive vulnerabilities. This environment builds on the Cyber Operations Research Gym (CybORG)\cite{standen2021cyborg} and models APTs using a Hidden Markov Model (HMM).

\begin{figure}[htb]
    \centering
   \vspace{-2mm} \includegraphics[width=.46\textwidth]{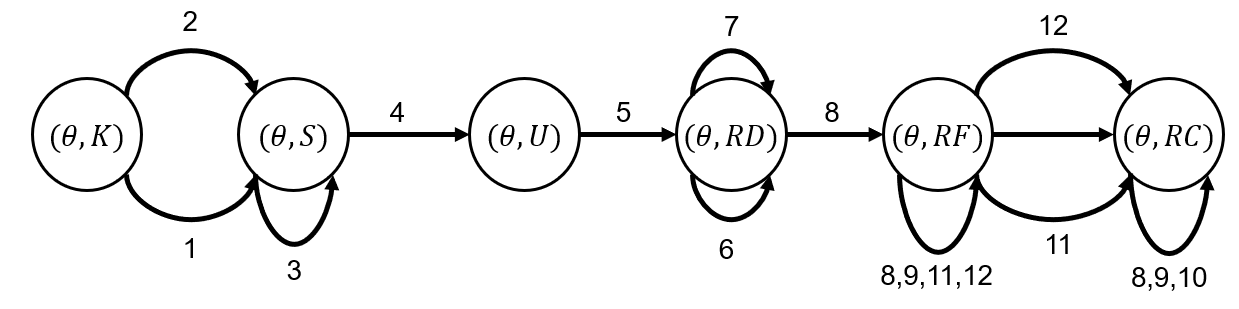}\vspace{-2mm}
    \caption{State transition diagram of a red agent's life stage in PsybORG$^+$. An APT life stage starts when the IP of a host is known. Actions 1-7 are actions in CybORG, while actions 8-12 are newly added. The R stage is divided into 3 sub-stages: RD, RF, and RC, with overlapping action spaces. RD represents that the root shell is successfully built. RF indicates that some crackable files have been found on this host. RC means that at least one credential file has been validated.
    }
    \label{new_fsm}\vspace{-5mm}
\end{figure}

PsybORG$^+$ consists of 3 teams of agents: red, blue, and green. Green agents simulate common user behaviors in the network. Red agents take actions to comprise green agents’ work, as shown in Figure \ref{new_fsm} and Table \ref{action_in_fsm}. Blue agents, acting as defenders, take the responsibility of preventing green agents from red agents’ attacks. 


\vspace{-3mm}\begin{table}[htb]
\centering
\caption{\label{action_in_fsm}Action Table in PsybORG$^+$}
\begin{tabular}{|c|c|c|}
	\hline Number&Action&time cost\\
        \hline 1&Aggressive service discovery&1\\
        \hline 2&Stealth service discovery&3\\
        \hline 3&Decoy detection&2\\
        \hline 4&Service exploit&4\\
        \hline 5&Privilege Escalate& 2\\
        \hline 6&Degrade service&2\\
        \hline 7&Impact(Stop OT service)&2\\
        \hline 8&Files discovery&1\\
        \hline 9&Bruteforce file cracking&3\\
        \hline 10&Password-based file cracking&1\\
        \hline 11&Credential file confirming&1\\
        \hline 12&Credential file disconfirming&1\\
    \hline
\end{tabular}
\end{table}

Files discovery is used to model the function of some automated reconnaissance tools, like 'DirBuster', which can scan and list files and directories on a host, providing attackers with an overview of the file system structure.

Files discovery can find all files' names, paths and values in the host. The hardness is not observed for the red agent. After calling files discovery, if there are files on this host, the state will transit from RD to RF, which means potential file targets are found on this host. Then, further actions can be taken. Files discovery can also be called to discover new files on the host.

Bruteforce file cracking is used to simulate file decryption and password cracking actions. Attackers attempt to gain unauthorized access to protected files by either doing brute force password enumeration. In PsybORG$^+$, brute force file cracking has a failure rate equal to the target file's hardness. 

Credential files, which contain filename-password mappings, can be found on the server. However, some credential files are decoys deployed by the defender to mislead attackers, which contain false filename-password mappings. These passwords can not help the attacker crack the file. The attacker can take actions to confirm or disconfirm a credential file. If red agents trust the credential file, they can do password-based file cracking to crack a file with a 100\% success rate.

Trigger is a system condition that can stimulate the attackers to take some actions revealing their cognitive vulnerability. Assuming there are some password-protected files with sounding filenames in the subnet, we can place some credential files as the trigger of the sunk cost fallacy. Once the attacker is exposed to those credential files, it would invest time and effort into cracking passwords and testing credential files.

\subsection{Biases state in PsybORG$^+$}
To illustrate the functionalities and capabilities of PsybORG$^+$, we  focus on the following 3 biases: loss aversion, sunk cost fallacy, and confirmation bias. We consider 2 levels for each bias: low and high, and hence $N$ is set to 3, and $|V_n|$ is set to 2 for each bias $n\in N$. An APT attacker's biases-influenced factor $\lambda \in \mathbb{R}^3$ can be represented by ($\lambda_l$,$\lambda_c$,$\lambda_s$). Table \ref{bias_group} lists the 8 biases state in PsybORG$^+$.

\vspace{-3mm}\begin{table}[htb]
\centering
\caption{\label{bias_group}Biases states}
\begin{tabular}{|c|c|c|c|}
	\hline Biases&Loss aversion&Confirmation bias&Sunk cost fallacy\\
        \hline $\theta_0$&Low&Low&Low \\
        \hline $\theta_1$&Low&Low&High \\
        \hline $\theta_2$&Low&High&Low \\
        \hline $\theta_3$&Low&High&High \\
        \hline $\theta_4$&High&Low&Low \\
        \hline $\theta_5$&High&Low&High \\
        \hline $\theta_6$&High&High&Low \\
        \hline $\theta_7$&High&High&High \\
    \hline
\end{tabular}\vspace{-0mm}
\end{table}

The expectation gain or loss of taking a service discovery is used to represent $\omega$. Both of $\rho$ and $\mu$ are set as 1. We can infer a red agent's loss aversion by analyzing the proportion of aggressive service discovery actions within the overall service discovery actions. 

$r(z)$ is the value of file $z$, and $c(z)$ is the times of  attempts the agent applies on $z$. We can also observe a red agent's sunk cost fallacy through the maximum number of file cracking attempts the agent applies on a particular file.

\section{Synthetic data generation}
Collecting a sufficient amount of attacker action data on real network systems can be challenging, as can constructing sufficiently diverse attack scenarios. Consequently, the analysis of attacker behavioral patterns can be often incomplete. We developed a classification model and a PsybORG$^+$-based simulator.
The classification model predicts APT attackers' cognitive biases based on their action sequences. The simulator uses these predictions to generate synthetic data by interacting with PsybORG$^+$.

\subsection{Experimental settings}
We built a dataset with 400 pieces of parameters (50 pieces of parameters for each biases state). 
Each subnet in PsybORG$^+$ has 3-10 user hosts and 1-6 server hosts. In the initialization part, 30 common files are generated on every host. The simulation step is 600 steps. There is a 0.1 probability of generating a credential file on each host, which contains passwords for 3-5 files. According to the central limit theorem, $p(\lambda|\theta)$ in the dataset follows the Gaussian distribution, as shown in Table \ref{bias_distribution}. The simulator uses these learned estimated distributions to sample $\lambda_l$, $\lambda_c$, and $\lambda_s$ for any inputted biases state.



  \vspace{-3mm} \begin{table}[htb]
\centering
\caption{\label{bias_distribution}Estimated parameter distribution}
\begin{tabular}{|c|c|c|c|}
	\hline Biases&p($\lambda_l|\theta$)&p($\lambda_c|\theta$)&p($\lambda_s|\theta$)\\
        \hline $\theta_0$&$N(0.5, 0.04)$&$N(0.19, 0.01)$&$N(201, 1764)$ \\
        \hline $\theta_1$&$N(0.5, 0.04)$&$N(0.19, 0.01)$&$N(798, 1521)$ \\
        \hline $\theta_2$&$N(0.5, 0.04)$&$N(0.79, 0.01)$&$N(201, 1764)$ \\
        \hline $\theta_3$&$N(0.5, 0.04)$&$N(0.79, 0.01)$&$N(798, 1521)$ \\
        \hline $\theta_4$&$N(1.51, 0.04)$&$N(0.19, 0.01)$&$N(201, 1764)$\\
        \hline $\theta_5$&$N(1.51, 0.04)$&$N(0.19, 0.01)$&$N(798, 1521)$\\
        \hline $\theta_6$&$N(1.51, 0.04)$&$N(0.79, 0.01)$&$N(201, 1764)$\\
        \hline $\theta_7$&$N(1.51, 0.04)$&$N(0.79, 0.01)$&$N(798, 1521)$\\
    \hline
\end{tabular}\vspace{-5mm}
\end{table}

\subsection{Biases state inference}
\subsubsection{Bayesian inference algorithm} Assuming p($\lambda|\theta$) listed in Table \ref{bias_distribution} and initial biases distribution p($\theta$) are known, we can use the Bayesian inference algorithm to do biases state inference for confirmation bias and loss aversion.
Since attackers with each biases state $\theta$ account for an equal portion of the dataset, $p(\theta)$ is set to 0.125.
To facilitate the discussion, we introduce the following notations: $u_a$ denotes taking aggressive service discovery; $u_s$ represents taking stealth service discovery; $u_c$ denotes taking credential file confirming action; $u_d$ represents finding disconfirming evidence for a credential file. We have $p(u_a|\lambda)$ = $\gamma$($a$,$s$,$\lambda_l$) in (1), $p(u_s|\lambda)$ = 1 - $\gamma$($a$,$s$), $p(u_c)$=$\lambda_c$, and $p(u_d)$=1-$\lambda_c$. Therefore, at time $t$, the observed attacker action $u_t\in\{u_a, u_c, u_s, u_d\}$ is given by $p(u_t|\theta)=\int p(u_t|\lambda)p(\lambda|\theta) \, d\lambda$ can be computed by numerical integration, as shown in Table \ref{action_probability}. 

The experimental results show that the Bayesian inference algorithm achieves an accuracy rate of 0.965 in inferring the biases state $\theta$ given the action sequence $\mathbf{u}$. Additionally, the average Cross Entropy for estimating $p(\mathbf{u}|\theta)$ is 0.038. 

However, the Bayesian inference can not infer the sunk cost fallacy, because files' value and hardness can also influence choice of file cracking target. We need the data-driven classification model to infer the sunk cost fallacy bias.

  \vspace{-3mm} \begin{table}[htb]
\centering
\caption{\label{action_probability}Emission probability in Bayesian inference}
\begin{tabular}{|c|c|c|c|c|}
	\hline Biases&$p(u_a|\theta)$&$p(u_s|\theta)$&$p(u_c|\theta)$&$p(u_d|\theta)$\\
        \hline $\theta_0$&0.66&0.34&0.19&0.81 \\
        \hline $\theta_1$&0.66&0.34&0.19&0.81 \\
        \hline $\theta_2$&0.66&0.34&0.79&0.21 \\
        \hline $\theta_3$&0.66&0.34&0.79&0.21 \\
        \hline $\theta_4$&0.33&0.67&0.19&0.81 \\
        \hline $\theta_5$&0.33&0.67&0.19&0.81 \\
        \hline $\theta_6$&0.33&0.67&0.79&0.21 \\
        \hline $\theta_7$&0.33&0.67&0.79&0.21 \\
    \hline
\end{tabular}\vspace{-3mm}
\end{table}





\subsubsection{Data-driven classification model}
\begin{figure}[htb]
    \centering
   \vspace{-3mm} \includegraphics[width=0.43\textwidth]{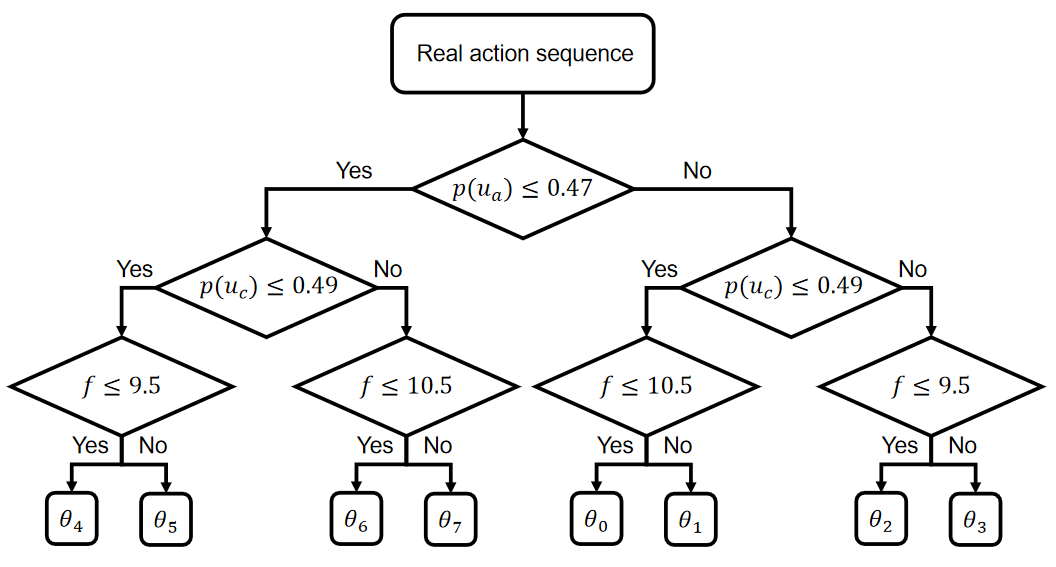}
      \vspace{-3mm} \caption{Data-driven Decision Tree. $p(u_a)$ denotes the rate of taking aggressive service discovery. $p(u_c)$ represents the rate of doing credential file confirming. $f$ represents the maximum attempts of file cracking applied to the same file in the action sequence.}
    \label{decision_tree_non_trigger}\vspace{-3mm}
\end{figure}

There is a decision-tree based classification model in PsybORG$^+$ to do biases state inference. The data metric learned by the model is presented in Figure \ref{decision_tree_non_trigger}. The model achieves an accuracy rate of 0.95 in the classification of loss aversion and a 0.99 accuracy rate on confirmation bias, which is similar to the performance of Bayesian inference algorithm. However, it only has an accuracy rate of 0.83 on sunk cost fallacy bias classification. That's might because the value and hardness of each file would also influence the choosing of target in File cracking action.

We evaluate the simulator by assessing the similarity between real action sequences and synthetic action sequences generated by sampled parameters. The results of red agents with random parameters and those with real parameters are set as baselines for assessing the performance of our simulator.

\begin{figure*}[htb] 
	\centering  
	\subfigtopskip=2pt
	\subfigbottomskip=2pt 
	\subfigcapskip=-5pt
	\subfigure[Rate of doing aggressive service discovery]{
		\label{aggressive_distance}
		\includegraphics[width=0.3\linewidth]{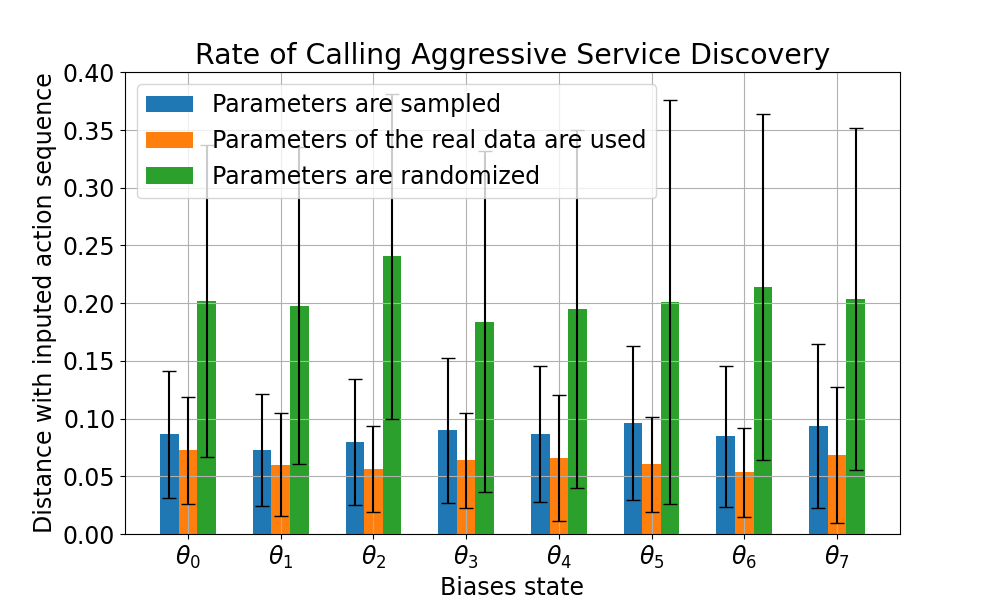}}
        \quad
	\subfigure[Rate of doing confirming credential file checking]{
		\label{confirm_evaluation}
		\includegraphics[width=0.3\linewidth]{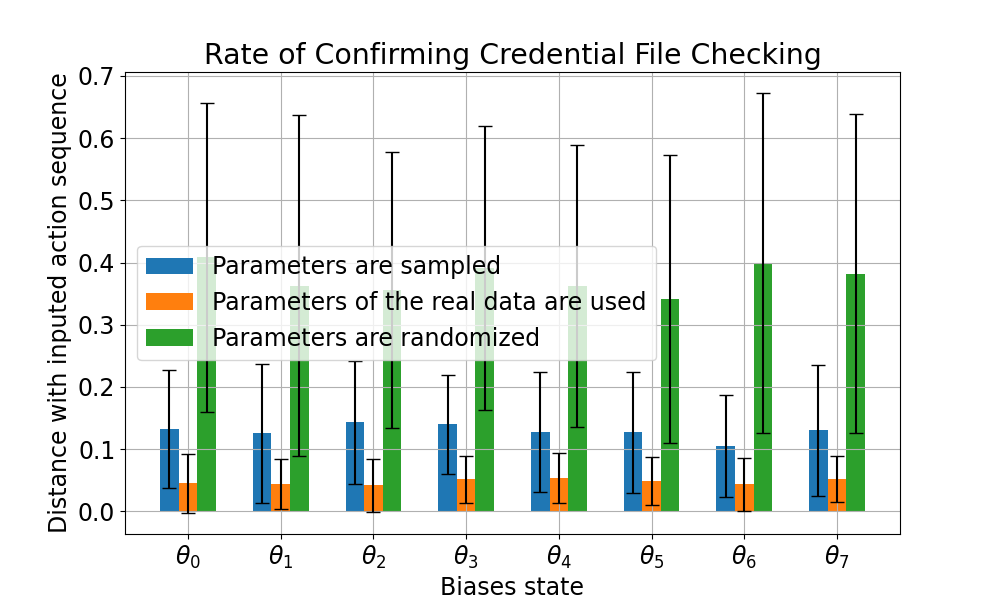}}
        \quad
        \subfigure[Maximum attempts of cracking file on the same file]{
		\label{sunk_cost_fallacy_evaluation}
		\includegraphics[width=0.3\linewidth]{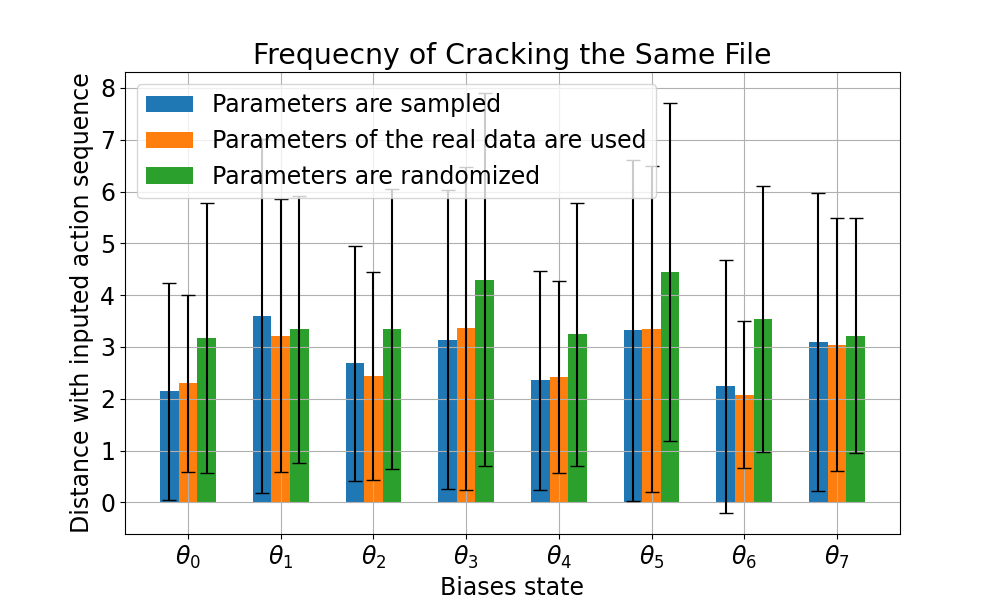}}
	\caption{Evaluation of synthetic data. An attacker's biases can be reflected in several characteristics: service discovery attempts, credential file checking, and target selection for cracking files. Therefore, we use these features as metrics to evaluate the quality of the synthetic data. The real parameters can serve as an upper bound, while random parameters can be considered as a lower bound for evaluating the simulator's effectiveness.}
	\label{compare2}\vspace{-5mm}
\end{figure*}

As shown in Figure \ref{compare2} and Table \ref{avg_std}, our simulator significantly outperforms the random algorithm in the service discovery and credential file checking simulation. However, for file cracking behaviors, the average distances among the three groups of parameters are similar, and all parameters exhibit high standard deviations. This indicates that PsybORG$^+$ is not effective in modeling the sunk cost fallacy. 

\begin{table}[htb]
    \centering
    \vspace{-4mm}
    \caption{Distance between synthetic data and real data\label{avg_std}}
    \begin{tabular}{ccccccc}
        \toprule
        &{Biases} & {Sampled Param.} & {Real Param.} &{Random Param.} \\
        \midrule
\multirow{8}{*}{\rotatebox{90}{\textbf{Service discovery}}}
        & $\theta_0$ & 0.09 $\pm$ 0.05 & 0.07 $\pm$ 0.05 & 0.20 $\pm$ 0.13 \\
        & $\theta_1$ & 0.07 $\pm$ 0.05 & 0.06 $\pm$ 0.04 & 0.20 $\pm$ 0.14 \\
        & $\theta_2$ & 0.08 $\pm$ 0.05 & 0.06 $\pm$ 0.04 & 0.24 $\pm$ 0.14 \\
        & $\theta_3$ & 0.09 $\pm$ 0.06 & 0.06 $\pm$ 0.04 & 0.18 $\pm$ 0.15 \\
        & $\theta_4$ & 0.09 $\pm$ 0.06 & 0.07 $\pm$ 0.05 & 0.19 $\pm$ 0.16 \\
        & $\theta_5$ & 0.10 $\pm$ 0.07 & 0.06 $\pm$ 0.04 & 0.20 $\pm$ 0.17 \\
        & $\theta_6$ & 0.08 $\pm$ 0.06 & 0.05 $\pm$ 0.04 & 0.21 $\pm$ 0.15 \\
        & $\theta_7$ & 0.09 $\pm$ 0.07 & 0.07 $\pm$ 0.06 & 0.20 $\pm$ 0.15 \\
        \midrule
        \multirow{8}{*}{\rotatebox{90}{\textbf{Cred file checking}}}
        & $\theta_0$ & 0.13 $\pm$ 0.10 & 0.04 $\pm$ 0.05 & 0.41 $\pm$ 0.25 \\
        & $\theta_1$ & 0.13 $\pm$ 0.11 & 0.04 $\pm$ 0.04 & 0.36 $\pm$ 0.27 \\
        & $\theta_2$ & 0.14 $\pm$ 0.10 & 0.04 $\pm$ 0.04 & 0.36 $\pm$ 0.22 \\
        & $\theta_3$ & 0.14 $\pm$ 0.08 & 0.05 $\pm$ 0.04 & 0.39 $\pm$ 0.23 \\
        & $\theta_4$ & 0.13 $\pm$ 0.10 & 0.05 $\pm$ 0.04 & 0.36 $\pm$ 0.23 \\
        & $\theta_5$ & 0.13 $\pm$ 0.10 & 0.05 $\pm$ 0.04 & 0.34 $\pm$ 0.23 \\
        & $\theta_6$ & 0.10 $\pm$ 0.08 & 0.04 $\pm$ 0.04 & 0.40 $\pm$ 0.27 \\
        & $\theta_7$ & 0.13 $\pm$ 0.11 & 0.05 $\pm$ 0.04 & 0.38 $\pm$ 0.26 \\
        \midrule
        \multirow{8}{*}{\rotatebox{90}{\textbf{File cracking}}}
        & $\theta_0$ & 2.14 $\pm$ 2.10 & 2.30 $\pm$ 1.71 & 3.18 $\pm$ 2.61 \\
        & $\theta_1$ & 3.60 $\pm$ 3.41 & 3.22 $\pm$ 2.64 & 3.34 $\pm$ 2.57 \\
        & $\theta_2$ & 2.68 $\pm$ 2.27 & 2.44 $\pm$ 2.01 & 3.34 $\pm$ 2.70 \\
        & $\theta_3$ & 3.14 $\pm$ 2.89 & 3.36 $\pm$ 3.12 & 4.30 $\pm$ 3.60 \\
        & $\theta_4$ & 2.36 $\pm$ 2.11 & 2.42 $\pm$ 1.86 & 3.24 $\pm$ 2.53 \\
        & $\theta_5$ & 3.32 $\pm$ 3.29 & 3.34 $\pm$ 3.15 & 4.44 $\pm$ 3.26 \\
        & $\theta_6$ & 2.24 $\pm$ 2.45 & 2.08 $\pm$ 1.43 & 3.54 $\pm$ 2.57 \\
        & $\theta_7$ & 3.10 $\pm$ 2.87 & 3.04 $\pm$ 2.44 & 3.22 $\pm$ 2.27 \\
        \bottomrule
    \end{tabular}\vspace{-4mm}
\end{table}

\section{Conclusion}
In this work, we have developed a mathematical model of APT attackers incorporating base rate neglect, loss aversion, confirmation bias, and the sunk cost fallacy. This model has been integrated into an APT simulation environment to create PsybORG$^+$, a multi-agent cybersecurity simulation platform designed to trigger and detect cognitive biases in attackers and simulate their behaviors. We have evaluated the performance of PsybORG$^+$ through a series of experiments, which demonstrated its effectiveness in simulating APT attack behaviors. The simulator enables the generation of synthetic data, aligns with human subject research data, and facilitates the design of defense mechanisms. PsybORG$^+$ is poised to play a critical role in benchmarking cyberpsychology studies and advancing research in this field.

\section{Acknowledgement}

This research is based upon work supported in part by the Office of the Director of National Intelligence (ODNI), Intelligence Advanced Research Projects Activity (IARPA) under Reimagining Security with Cyberpsychology-Informed Network Defenses (ReSCIND) program contract N66001-24-C-4504. The views and conclusions contained herein are those of the authors and should not be interpreted as necessarily representing the official policies, either expressed or implied, of ODNI, IARPA, or the U.S. Government. The U.S. Government is authorized to reproduce and distribute reprints for governmental purposes notwithstanding any copyright annotation therein.

\bibliographystyle{IEEEtran} 

\bibliography{paper}

\begin{thebibliography}{10}
\providecommand{\url}[1]{#1}
\csname url@samestyle\endcsname
\providecommand{\newblock}{\relax}
\providecommand{\bibinfo}[2]{#2}
\providecommand{\BIBentrySTDinterwordspacing}{\spaceskip=0pt\relax}
\providecommand{\BIBentryALTinterwordstretchfactor}{4}
\providecommand{\BIBentryALTinterwordspacing}{\spaceskip=\fontdimen2\font plus
\BIBentryALTinterwordstretchfactor\fontdimen3\font minus \fontdimen4\font\relax}
\providecommand{\BIBforeignlanguage}[2]{{%
\expandafter\ifx\csname l@#1\endcsname\relax
\typeout{** WARNING: IEEEtran.bst: No hyphenation pattern has been}%
\typeout{** loaded for the language `#1'. Using the pattern for}%
\typeout{** the default language instead.}%
\else
\language=\csname l@#1\endcsname
\fi
#2}}
\providecommand{\BIBdecl}{\relax}
\BIBdecl

\bibitem{chen2014study}
P.~Chen, L.~Desmet, and C.~Huygens, ``A study on advanced persistent threats,'' in \emph{Communications and Multimedia Security: 15th IFIP TC 6/TC 11 International Conference, CMS 2014, Aveiro, Portugal, September 25-26, 2014. Proceedings 15}.\hskip 1em plus 0.5em minus 0.4em\relax Springer, 2014, pp. 63--72.

\bibitem{strom2018mitre}
B.~E. Strom, A.~Applebaum, D.~P. Miller, K.~C. Nickels, A.~G. Pennington, and C.~B. Thomas, ``Mitre att\&ck: Design and philosophy,'' in \emph{Technical report}.\hskip 1em plus 0.5em minus 0.4em\relax The MITRE Corporation, 2018.

\bibitem{fahad2023securing}
M.~Fahad, H.~Airf, A.~Kumar, and H.~K. Hussain, ``Securing against apts: Advancements in detection and mitigation,'' \emph{BIN: Bulletin Of Informatics}, vol.~1, no.~2, 2023.

\bibitem{zhu2018multi}
Q.~Zhu and S.~Rass, ``On multi-phase and multi-stage game-theoretic modeling of advanced persistent threats,'' \emph{IEEE Access}, vol.~6, pp. 13\,958--13\,971, 2018.

\bibitem{huang2018analysis}
L.~Huang and Q.~Zhu, ``Analysis and computation of adaptive defense strategies against advanced persistent threats for cyber-physical systems,'' in \emph{Decision and Game Theory for Security: 9th International Conference, GameSec 2018, Seattle, WA, USA, October 29--31, 2018, Proceedings 9}.\hskip 1em plus 0.5em minus 0.4em\relax Springer, 2018, pp. 205--226.

\bibitem{zhu2013game}
Q.~Zhu and T.~Ba{\c{s}}ar, ``Game-theoretic approach to feedback-driven multi-stage moving target defense,'' in \emph{International conference on decision and game theory for security}.\hskip 1em plus 0.5em minus 0.4em\relax Springer, 2013, pp. 246--263.

\bibitem{huang2020dynamic}
L.~Huang and Q.~Zhu, ``A dynamic games approach to proactive defense strategies against advanced persistent threats in cyber-physical systems,'' \emph{Computers \& Security}, vol.~89, p. 101660, 2020.

\bibitem{han2020unicorn}
X.~Han, T.~Pasquier, A.~Bates, J.~Mickens, and M.~Seltzer, ``Unicorn: Runtime provenance-based detector for advanced persistent threats,'' \emph{arXiv preprint arXiv:2001.01525}, 2020.

\bibitem{alrehaili2021hybrid}
M.~Alrehaili, A.~Alshamrani, and A.~Eshmawi, ``A hybrid deep learning approach for advanced persistent threat attack detection,'' in \emph{Proceedings of the 5th International Conference on Future Networks and Distributed Systems}, 2021, pp. 78--86.

\bibitem{eke2019use}
H.~N. Eke, A.~Petrovski, and H.~Ahriz, ``The use of machine learning algorithms for detecting advanced persistent threats,'' in \emph{Proceedings of the 12th international conference on security of information and networks}, 2019, pp. 1--8.

\bibitem{joloudari2020early}
J.~H. Joloudari, M.~Haderbadi, A.~Mashmool, M.~GhasemiGol, S.~S. Band, and A.~Mosavi, ``Early detection of the advanced persistent threat attack using performance analysis of deep learning,'' \emph{IEEE Access}, vol.~8, pp. 186\,125--186\,137, 2020.

\bibitem{kahneman1984choices}
D.~Kahneman and A.~Tversky, ``Choices, values, and frames.'' \emph{American psychologist}, vol.~39, no.~4, p. 341, 1984.

\bibitem{tversky1981framing}
A.~Tversky and D.~Kahneman, ``The framing of decisions and the psychology of choice,'' \emph{science}, vol. 211, no. 4481, pp. 453--458, 1981.

\bibitem{tversky1988contingent}
A.~Tversky, S.~Sattath, and P.~Slovic, ``Contingent weighting in judgment and choice.'' \emph{Psychological review}, vol.~95, no.~3, p. 371, 1988.

\bibitem{lemay2018cognitive}
A.~Lemay and S.~Leblanc, ``Cognitive biases in cyber decision-making,'' in \emph{Proceedings of the 13th International Conference on Cyber Warfare and Security}, 2018, p. 395.

\bibitem{ferguson2021examining}
K.~J. Ferguson-Walter, M.~M. Major, C.~K. Johnson, and D.~H. Muhleman, ``Examining the efficacy of decoy-based and psychological cyber deception,'' in \emph{30th USENIX security symposium (USENIX Security 21)}, 2021, pp. 1127--1144.

\bibitem{kahneman1973psychology}
D.~Kahneman and A.~Tversky, ``On the psychology of prediction.'' \emph{Psychological review}, vol.~80, no.~4, p. 237, 1973.

\bibitem{nickerson1998confirmation}
R.~S. Nickerson, ``Confirmation bias: A ubiquitous phenomenon in many guises,'' \emph{Review of general psychology}, vol.~2, no.~2, pp. 175--220, 1998.

\bibitem{schmidt2005loss}
U.~Schmidt and H.~Zank, ``What is loss aversion?'' \emph{Journal of risk and uncertainty}, vol.~30, pp. 157--167, 2005.

\bibitem{kahneman1979prospect}
D.~Kahneman and A.~Tversky, ``Prospect theory - analysis of decision under risk,'' \emph{Econometrica}, vol.~47, no.~2, pp. 263--291, 1979.

\bibitem{friedman2007searching}
D.~Friedman, K.~Pommerenke, R.~Lukose, G.~Milam, and B.~A. Huberman, ``Searching for the sunk cost fallacy,'' \emph{Experimental Economics}, vol.~10, pp. 79--104, 2007.

\bibitem{standen2021cyborg}
M.~Standen, M.~Lucas, D.~Bowman, T.~J. Richer, J.~Kim, and D.~Marriott, ``Cyborg: A gym for the development of autonomous cyber agents,'' \emph{arXiv preprint arXiv:2108.09118}, 2021.

\end{thebibliography}
\end{document}